\documentclass[prb,preprint]{revtex4}

\usepackage{amsmath}  
\usepackage{amsfonts} 
\usepackage{amssymb}
\usepackage{graphicx,color}
\usepackage{epstopdf}

\definecolor{r}{rgb}{1,0,0}   
\definecolor{g}{rgb}{0,1,0}   
\definecolor{b}{rgb}{0,0,1}
\begin{document}
\title{Stretching Rubber, Stretching Minds:\\a polymer physics lab for teaching entropy}
\author{Theodore A. Brzinski III}
\email{tbrzins@ncsu.edu}
\author{Karen E. Daniels}
\affiliation{Department of Physics, NC State University, 2401 Stinson Drive, Raleigh, NC 27705-8202, USA}

\date{\today}

\begin{abstract}
Entropy is a difficult concept to teach using real-world examples. Unlike temperature, pressure, volume, or work, it's not a quantity which most students encounter in their day-to-day lives. Even the way entropy is often qualitatively described, as a measure of disorder, is incomplete and can be misleading. In an effort to address these obstacles, we have developed a simple laboratory activity, the stretching of an elastic rubber sheet, intended to give students hands-on experience with the concepts of entropy, temperature and work in both adiabatic and quasistatic processes.  We present two versions of the apparatus: a double-lever system, which may be reproduced with relatively little cost, and a commercial materials testing system, which provides students experience with scientific instrumentation that is used in research.
\end{abstract}
\maketitle

\section{Introduction}

Rubber has long been proposed as a subject for discussion~\cite{Information2002,Smith2002,Pellicer2001,Marx2000,Nash1979,Sethna06statisticalmechanics, schroeder1999introduction}, and as a useful demonstration~\cite{Galley2007,Dole1977,Mullen1975,Brown1963} or laboratory excercise~\cite{Ritacco2014,Roundy2013,Euler2008,Byrne2007,Mitschele1997,Savarino1991a,Bader1981,Mark1981,Carroll1963} in undergraduate thermodynamics curriculum. Rubber is an attractive choice for undergraduate education because simple and tractable thermodynamic models, using little more than the First and Second Laws of Thermodynamics, can be powerfully predictive. Furthermore demonstrations or lab activities utilizing rubber are generally affordable and tactile, and given student's familiarity with rubber in their daily lives (tires, rubber bands, etc.), the subject is both relevant and memorable.

Rubber's thermodynamic phenomenology is also often counter-intuitive. For example, the Gough-Joule effect: rubber that is under tension will exhibit a negative thermal expansion coefficient. Another surprising phenomenon is that a rapidly stretched rubber band will heat up, while a stretched band that is suddenly released will cool to below the ambient temperature. These traits challenge student preconceptions, and lead them to think more deeply about the concepts involved: entropy, temperature, work, heat and the Laws of Thermodynamics.

\begin{figure}
\caption{A schematic of the simple model employed by students in the analysis of the laboratory exercises we describe here. A polymer is modeled as a 1-D chain of segments which contribute $\pm a$ to the total polymer length $L$.}
\label{fig:Model}\includegraphics[width=0.9\linewidth]{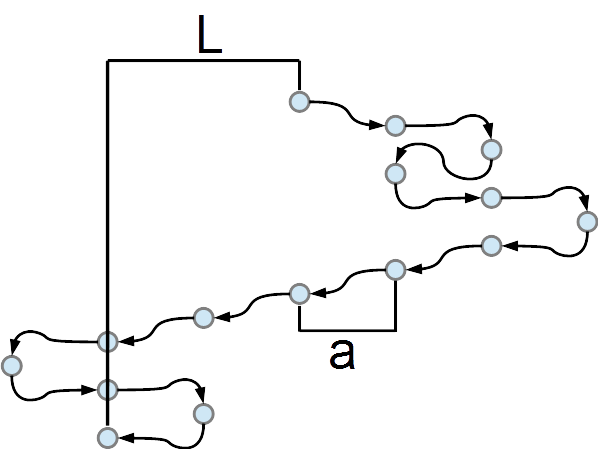}
\end{figure}

These phenomena defy intuition, yet students can explain them with a simple, 1-D model, with math accessible to undergraduate students. Rubber is composed of polymers: long chains of repeating molecular structures (monomers) which are free to bend and twist.  The students model these monomers as fixed-length segments which contribute length $\pm  a$ to the overall length $L$ of the polymer, as pictured in Fig.~\ref{fig:Model}. This cartoon model allows students to relate the macrostate of the system to the conformation of the constituent molecules, and by doing so, understand rubber's surprising thermomechanics.

While numerous instructional labs have explored the Gough-Joule effect~\cite{Ritacco2014,Roundy2013,Byrne2007,Mitschele1997,Savarino1991a,Bader1981,Mark1981,Carroll1963}, they universally characterize the system via its statics, and the procedures emphasize the equation of state rather than the microscopic origins of the phenomenon. Rubber's thermal response to  adiabatic processes, on the other hand, seems to have previously been reserved for lecture demonstration~\cite{Brown1963}.  We present a laboratory exercise which explores these behaviors in two parts: First, students measure the change in temperature of rubber in response to rapid stretching. Second, starting with stretched rubber, students alter the rubber's temperature, and measure the resulting change in tension. In both cases, students are asked to compare their observations of the thermal and mechanical dynamics to analytic predictions and explain discrepancies in terms of their assumptions in writing the microscopic model. Two versions of the apparatus are presented: a double lever and a commercial materials testing system.

\section{Polymer theory}

While the Gough-Joule effect challenges student expectations, it is simple to explain in the context of the cartoon model introduced above.  As the temperature of rubber is increased, the molecules wriggle about more energetically, making the more entropic (and thus shorter) configuration favorable.  This explanation can be arrived at analytically as follows: Assuming the entropy, $S$, is a natural function of polymer extension $L$ and the internal energy of the system, $U$, we can write:

\begin{equation}
    \label{eq:PartS}
	dS=\left(\frac{\partial S}{\partial L}\right)_{U}{dL}+\left(\frac{\partial S}{\partial U}\right)_{L}{dU}
\end{equation}
If we define the work done by the system as $\left\langle F\right\rangle{dL}$ where $F$ is the tension exerted on the polymer and the average is over $L$, and the heat added to the system as $T{dS}$, where $T$ is the rubber's temperature, we can write the First Law of Thermodynamics as:

\begin{equation}
	\label{eq:FirstL}
	{dU}=T{dS}+\left\langle F\right\rangle{dL}
\end{equation}
Combining Eqs.~(\ref{eq:PartS}) and (\ref{eq:FirstL}),  we can write
\begin{equation}
\left(\frac{\partial S}{\partial L}\right)_{U}{dL}+\left(\frac{\partial S}{\partial U}\right)_{L}{dU}=\frac{1}{T}{dU}-\frac{\left\langle F\right\rangle}{T}{dL}
\end{equation}
Thus, the tension exerted by the rubber is
\begin{equation}
	\label{eq:avgF}
	\left\langle F\right\rangle=-T\left(\frac{\partial S}{\partial L}\right)_{U}
\end{equation}

\begin{figure}
\caption{A simple cartoon contrasting a polymeric solid in both a stretched and relaxed state. Features like loops in the relaxed state are indicative of larger degeneracy.}
\label{fig:Cartoon1}\includegraphics[width=0.9\linewidth]{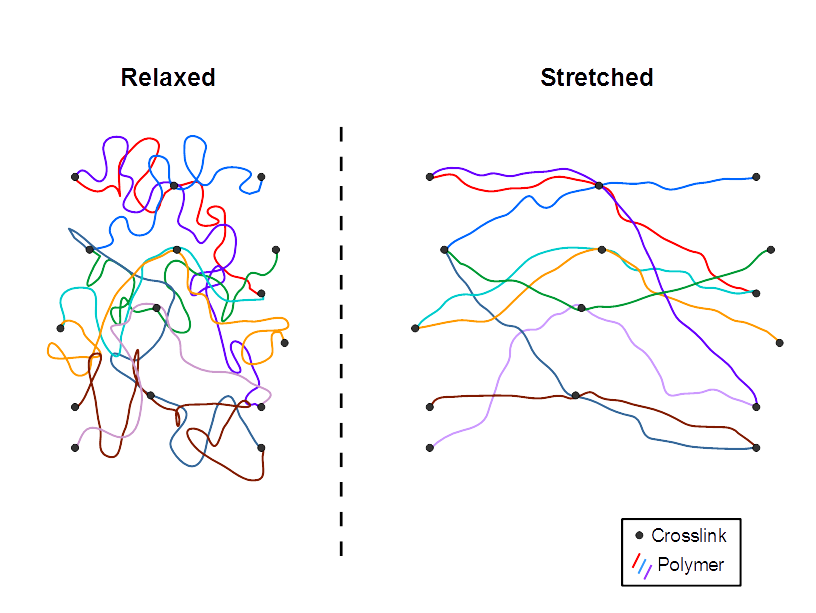}
\end{figure}

The polymers which make up rubber are connected at sites called 'crosslinks'~\cite{Jones02SoftMatter}. As depicted in Fig.~\ref{fig:Cartoon1}, when the rubber is relaxed, the cross-links will be close together and the polymers free to take any number of folded and looped configurations; However, when the rubber is stretched, and the crosslink separation approaches the length of the polymers, the polymers are pulled taut and are able to explore fewer configurations. The configurational entropy of the system is a measure of the polymers' freedom of motion, so as the rubber is stretched and the number of molecular states available becomes smaller, the entropy of the material is reduced. Therefore  $\left(\frac{\partial S}{\partial L}\right)_{U}$  must be negative, so according to Eq.~(\ref{eq:avgF}), the tension increases with temperature. Thus, as the temperature is increased, so long as the load on the rubber remains constant, it will contract. The precise form of  $\left(\frac{\partial S}{\partial L}\right)_{U}$ can be calculated from our cartoon model by counting the number of configurations that can produce a given value of $L$, reminiscent of polarization in a 1-D Ising model. For small extensions, by applying the binomial expansion, students can show that this model predicts the tension is linear in $L$, like Hooke's law.

Rubber's tendency to change temperature in response to being stretched or released is another exciting consequence of the molecular origins of rubber's elasticity. Rapidly stretching the rubber is an adiabatic process: The system can not exchange heat with the environment as fast as work is done on it, and the temperature will increase from $T_{o}$ to $T_{o}+\Delta{T}$. This is because the energy associated with the lost configurational entropy, $Q_{config}=T_{o}\Delta S$ assuming $\Delta{T}\lll{T_{o}}$, is converted to thermal vibrational energy. The vibrational energy of each polymer is that of $nN$ linear oscillators, $U_{vib}=nNkT$, where $n$ is the number of atoms that make up each of the $N$ monomers per polymer, so the heat capacity is $C=\frac{dU_{vib}}{dT}=nNk$. This gives the expected change in temperature:
\begin{equation}
\Delta{T}=\frac{Q_{config}}{C}=\frac{T_{o}\Delta{S}}{nNk}
\label{eq:dT}
\end{equation}

Thus, using the same 1-D model, students can calculate the change in the configurational entropy of a polymer as it is changed from one length to another, and also use that result to predict the magnitude of the temperature difference.  Because polymers consist of a known number of monomers of known chemical make-up, students can substitute real material values into their models and obtain dimensional results that can be tested directly against observation. The monomer in latex, isoprene, has chemical formula C$_5$H$_8$. Latex molecules can have any number of monomers, but for these exercises we suggest students use $N=20$.  

The derivations above are commonly included in thermal physics curricula, or are assigned as an exercise~\cite{Marx2000, Sethna06statisticalmechanics, schroeder1999introduction}.  This analysis requires students to employ many of the introductory concepts of thermodynamics and statistical mechanics. Particularly, the students explicitly utilize the First Law of Thermodynamics, consider heat and work, relate microscopic mechanics to bulk behavior, and utilize both the microscopic and bulk definitions of entropy. Furthermore, students must consider their assumptions, reason about which quantities are held constant, and interpret the physical meaning of their results. In discussing this analysis with students, it is helpful to point out that a carbon-carbon bond is approximately 1.5~angstroms in length, yet students will be looking for the effects of these atomic-scale predictions in a bulk sample measured in centimeters: understanding bulk behaviors as an ensemble effect of microscopic physics is key to the philosophy statistical mechanics.

\section{Experimental Apparatus}

Here we present two experimental apparatuses that have been used successfully for instructional laboratory modules for an advanced undergraduate thermal physics course for sophomores and juniors: a simple double lever, and a commercial Materials Testing System (MTS). Both apparatuses allow students to conduct the procedures described below, and separate instructions are provided for each apparatus. While the MTS offers better dynamic control of the extension, as well as higher resolution measurements, both apparatuses enable students to measure both the change in temperature due to adiabatic stretching of the rubber sheet, as well as to measure the tension as they perturb the rubber's temperature, and compare their results to the calculation of Marx et al.~\cite{Marx2000} and Eq.~\ref{eq:avgF} respectively. In our laboratory exercises, the sample studied is a sheet of silicone elastomer (Theraband brand), though any such elastic sheet or band would work.

All temperature measurements are taken at the surface of the rubber sheet with a contactless IR thermometer. The operation of an IR thermometer is itself an interesting application of thermodynamics, and can provide an interesting reading assignment in conjunction with these lab activities~\cite{Barron}.

\subsection{Commercial Materials Testing System}

\begin{figure}
\includegraphics[width=0.99\linewidth]{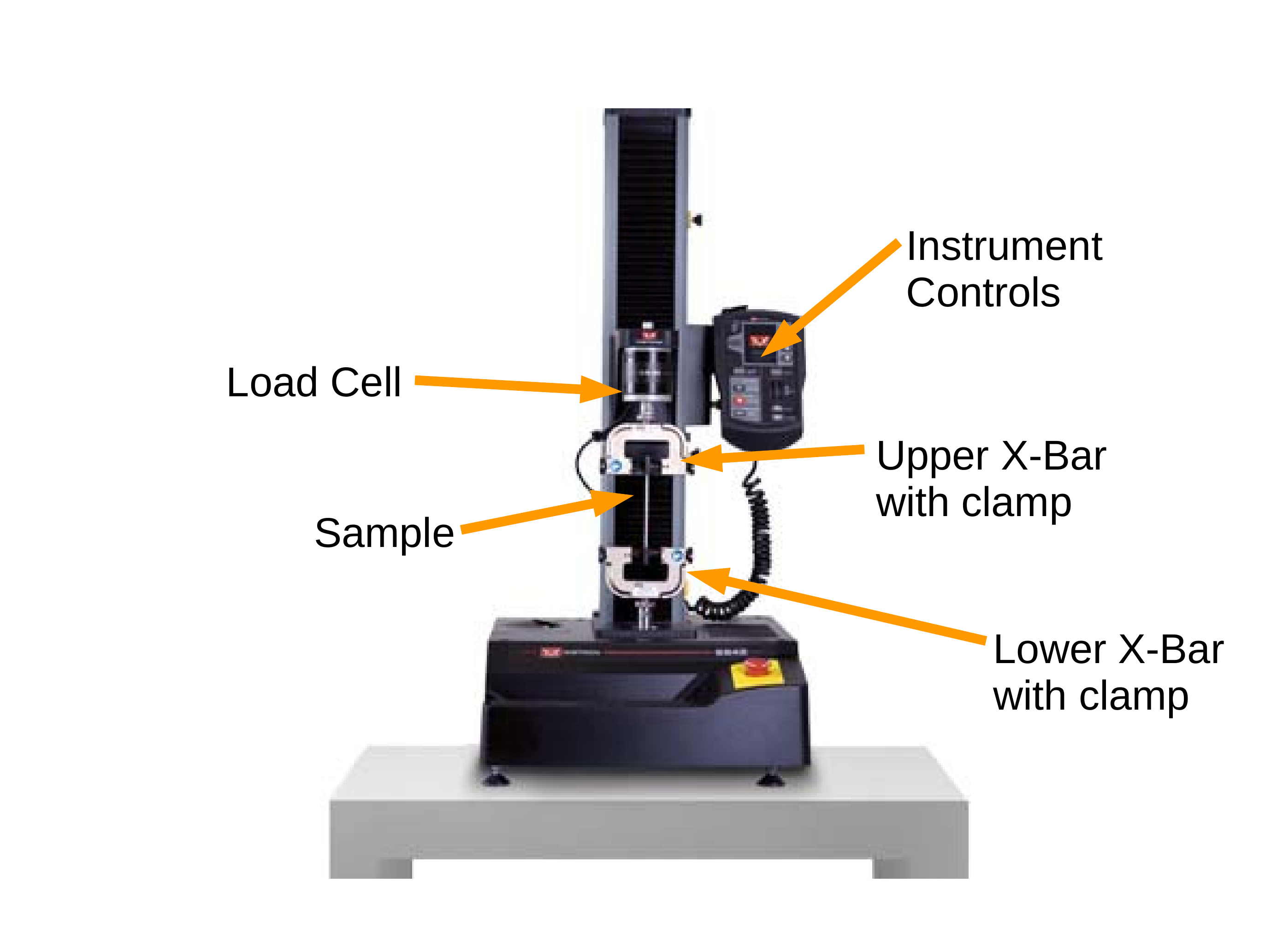}
\caption{A picture of the materials testing system (MTS) utilized for these experiments.  The key components are indicated with orange arrows and text labels in the image: the load cell, the upper and lower crossbars, the sample and the instrument controls.}
\label{fig:Instron}
\end{figure}

Materials Testing Systems are used extensively in research and industry to test the strength and response of materials under compression or tension.  The model we use (Instron model 5944) is pictured in Fig.~\ref{fig:Instron}. One crossbar (labeled as the upper crossbar in the figure) moves along a rigid rail, while a second is afixed to the apparatus base (labeled as the lower crossbar in the figure). Mounted to the upper crossbar is a high-resolution load cell. Finally, attached to both crossbars are clamps which hold the sample. In order to stretch the rubber in a controlled and uniform manner, a larger clamping surface is required, so the rubber is first clamped at both ends between pieces of t-slotted aluminum extrusion with applications of adhesive-backed sandpaper on the clamping surfaces.  The MTS then clamps to the extrusion rather than directly to the rubber. Most MTSs have the ability to pre-tension the sample, increasing the extension of the rubber until some (small) tension is achieved before re-zeroing both the tension and extension measurements.  Pretensioning in this way pulls the rubber taut without stretching it. Once the material is mounted and pretensioned in the MTS, the students can stretch the rubber at a prescribed rate to prescribed threshold values of either the extension or tension.

For many instructors,  securing access to an MTS may be difficult compared to the double lever described below. However, there are a few of significant advantages to utilizing an MTS. First, while both apparatuses record the tension at a sufficiently high data-rate, the MTS has the added benefits of simultaneously recording the extension and precisely controlling either the tension or the extension. This allows students to discuss and interpret the functional form of the relation between extension and tension while the double lever does not. Second, because commercial MTSs are primarily used in industrial and research settings, students are not only being exposed to an exciting, high quality piece of equipment, but are also learning applicable trade experience as they learn to use the system. Indeed, the brief experience students had with the MTS in completing this exercise led a small number to inquire about independent research projects which utilize the system. Finally, because most MTSs are designed to handle large batches for material testing, the software generally supports procedure files which prompt users to record any secondary measurements (like temperature, in the case of this lab), and ensure the correct test parameters and failsafes are used. Providing students with such files can help to reduce error, and demonstrate the power of the system to students without risking inadvertent damage or injury due to user error. That said, we would issue one final note of caution: these instruments present a crushing hazard if handled improperly, and can be damaged if the appropriate failsafes are not used.  As with any lab equipment, ensure students are informed of how to operate it safely. 

\subsection{Double Lever}

\begin{figure}
\includegraphics[width=0.85\linewidth]{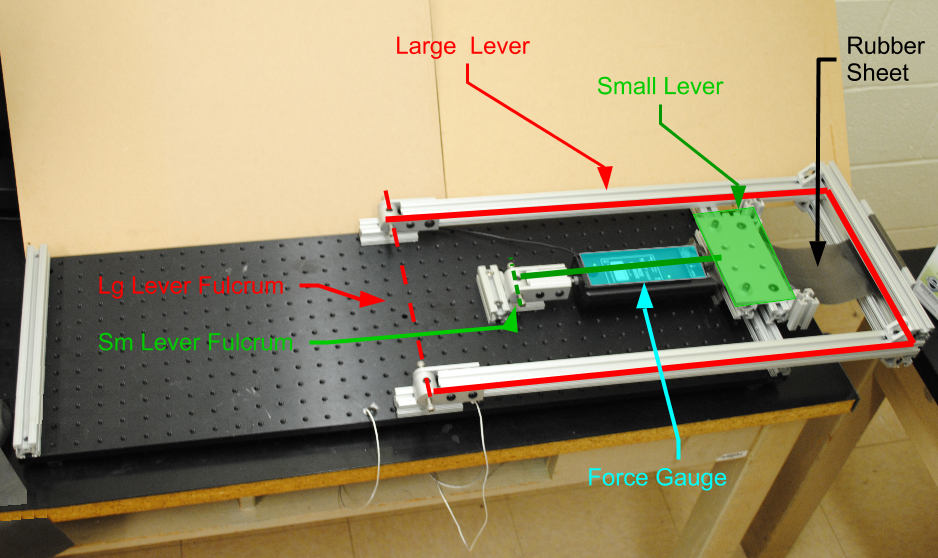}
\caption{A photograph of the first version of the double-lever apparatus. The larger lever is traced in red. As indicated, its fulcrum is along the center of the base-plate. The inner, small lever is highlighted in green, and rotates about an adjustable fulcrum which is offset from that of the larger, outer lever. The rubber is mounted between the two levers, and its tension is measured by a Pasco load cell (highlighted in cyan). For scale, the base plate has dimensions of 12~inches~x~36~inches.}
\label{fig:DblLever}
\end{figure}

A photograph of the double-lever apparatus is provided in Fig.~\ref{fig:DblLever}. A descriptive video can be found in the supplemental materials~\cite{SuppMat} . The key feature of  this apparatus is a pair of levers of different lengths whose fulcra share a plane, but are offset by some distance $d$. In Fig.~\ref{fig:DblLever}, the longer lever is highlighted in red, and the shorter in green. When both levers are oriented along the direction from the fulcrum of the longer lever towards that of the smaller lever  (as pictured), the distance between the ends of the levers are minimized. When the orientation of both levers is flipped 180$^\circ$, the distance between the ends of the levers increases by $2d$. The longer lever is anchored at the center of an optical breadboard which serves as the base-plate, and has threaded holes arrayed in a 1"-square grid. The shorter lever can be mounted to any of the rows of threaded holes, allowing for adjustments of $d$ in increments of 1~inch. The shorter lever incorporates a load cell (highlighted in cyan). The rubber sample is anchored between the load cell and the handle of the larger lever so that the load cell measures the tension exerted by the rubber.  The rubber is adjusted so that it is held taut, but not stretched. Therefore, when the lever orientations are reversed, the rubber is stretched by length $2d$. The lever enables students to produce considerable tension in a short time, thus approaching the adiabatic limit.

\section{Experimental Procedure}

The laboratory procedure is broken into two parts: A series of adiabatic extensions and releases during which students measure the changes in temperature associated with the cycle, and a 'sandbox' component during which the sample is held at a constant extension, and students are asked to perturb the sample temperature to observe the response and subsequent relaxation. Before undertaking the activity, students should have derived Eq.~\ref{eq:avgF} and completed the analytic prediction for $\Delta T$ during adiabatic stretching as outlined above. Example course documents for these exercises are included in our supplemental materials~\cite{SuppMat}.

\subsection{Preparation}

At the outset of the lab we find it helpful to encourage students to familiarize themselves with some of the equipment.  In particular, the infrared thermometer may require more nuanced operation than many students realize.  Specifically, most such thermometers provide a laser guide to indicate what surface is being measured. The laser is usually offset by about an inch from the center of the area of the actual measurement. Furthermore, the area measured increases as the square of the distance from the thermometer. Thus, we ask the students to practice by measuring several objects around the room, then to take 5 minutes or so to familiarize themselves with the operation and anatomy of the experimental apparatus, particularly if they are using the MTS. Finally, students are provided  with a scrap of rubber and a length of steel wire coil spring, and are asked to qualitatively compare and contrast the two, and to discuss what sets each item's spring constant. This final step helps students contextualize the experiment they are about to perform, and reminds them of the entropic, rather than atomic origins of the rubber's elasticity.

\subsection{Adiabatic Stretching}

\subsubsection{Material Testing System}
With the rubber in the unstretched position, students measure the length and temperature of the sample, and zero the force measurement. Next, they rapidly ($\geq 1$~cm/s) stretch the rubber to a predetermined length. Students should measure the temperature during the stretching process, and record the largest value observed, as the rubber will immediately begin exchanging heat with the environment, and quickly cool back toward room temperature.  Once it has been characterized in the stretched state, and returned to room temperature, the procedure is repeated in reverse, returning the rubber to the unstretched state, and generating a temperature drop. To ensure the results are reproducible, students repeat these measurements 3-5 times.

\subsubsection{Double Lever}
The procedure is largely the same as for the MTS: With the double lever in the unstretched position, students measure the length and temperature of the sample, and zero the force measurement.  Once they are ready to stretch the rubber, the students quickly swing the levers to the stretched position, and repeat their temperature and length measurements. As above, students should record the largest temperature observed, as the rubber will immediately begin to cool. Once the rubber's temperature has returned to roughly room temperature, students are instructed to quickly return the system to its unstretched state, and measure the resulting temperature and force drop.  To ensure the results are reproducible, students repeat these measurements 3-5 times.

\subsubsection{Analysis and discussion.}
The procedure for this component is straightforward, regardless of which experimental design is used, but also provides the most obvious comparison to theory: students have calculated changes in temperature. Do they match observation or not? Qualitatively, students should observe an agreement in the sign of the effect: adiabatic stretching leads to an increase in temperature, while adiabatic unstretching leads to a temperature drop. Quantitatively, students' estimates will generally be one or more orders of magnitude greater than the change measured, depending on their assumptions.

This is a useful opportunity to quell any impulse to label the experiment a failure, and instead ask students to identify and address the assumptions used in their predictions. Does using the appropriate number of dimensions improve or reduce the agreement between theory and observation?  In the calculation for Marx et. al~\cite{Marx2000} it is assumed that the polymer goes from unconstrained to fully extended. Is that valid?  What different assumptions might improve the theory? Did the entropy of the system change more or less than expected?

\begin{figure}
\includegraphics[width=0.9\linewidth]{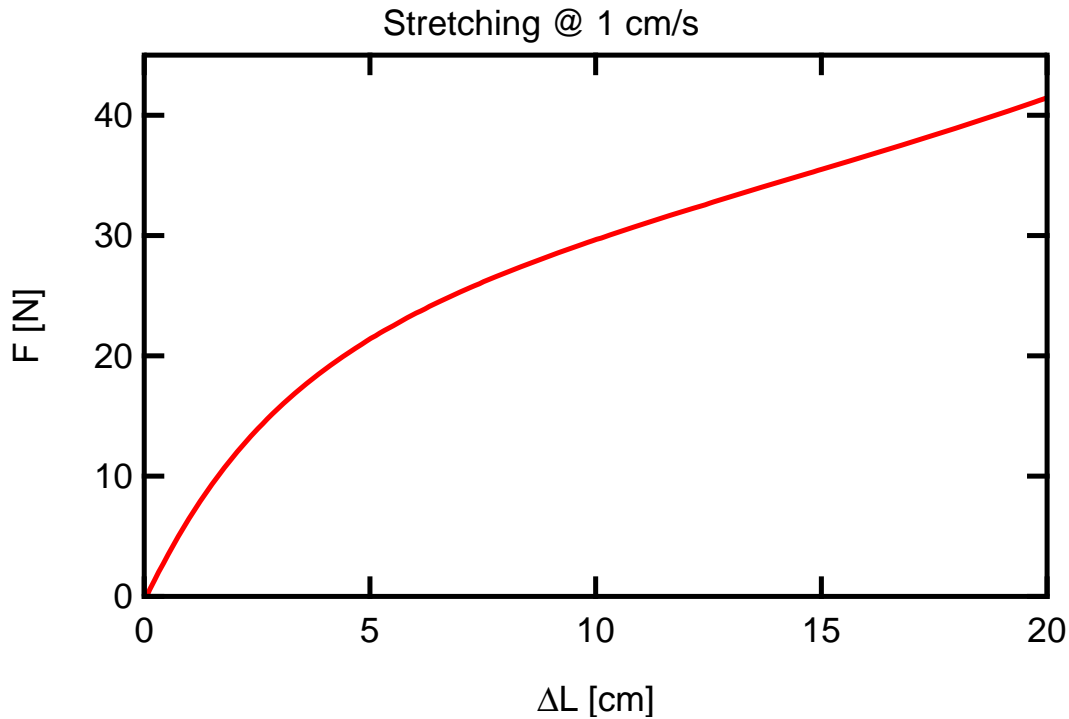}
\caption{Experimental values of tension plotted vs extension for a piece of rubber stretched adiabatically on the MTS.}
\label{fig:RapidStretch}
\end{figure}

We find that few students end up using their measurements of force or length in their discussion of $\Delta T$ unprompted, but with some guidance, students will note that, while the magnitude of $\Delta T$ is comparable whether stretching or unstretching, the change in force while the system returns to room temperature is not: the tension relaxes in the stretched state, but remains constant in the unstretched state.  This is consistent with Eq.~(\ref{eq:avgF}), which requires that that $\left(\frac{\partial S}{\partial L}\right)_{U}=0$ for an unstretched polymer. Is it consistent with the students' model for $S\left(L,U\right)$? Students working with an MTS may notice one more interesting feature: the tension of the band as a function of extension, as plotted in Fig.~\ref{fig:RapidStretch} doesn't match the linear form they derived for Eq.~(\ref{eq:avgF}). This is because the rubber's total internal energy and temperature are both changing during extension. 

\subsection{Force and Temperature}

\subsubsection{Material Testing System}
\begin{figure*}
\includegraphics[width=0.9\linewidth]{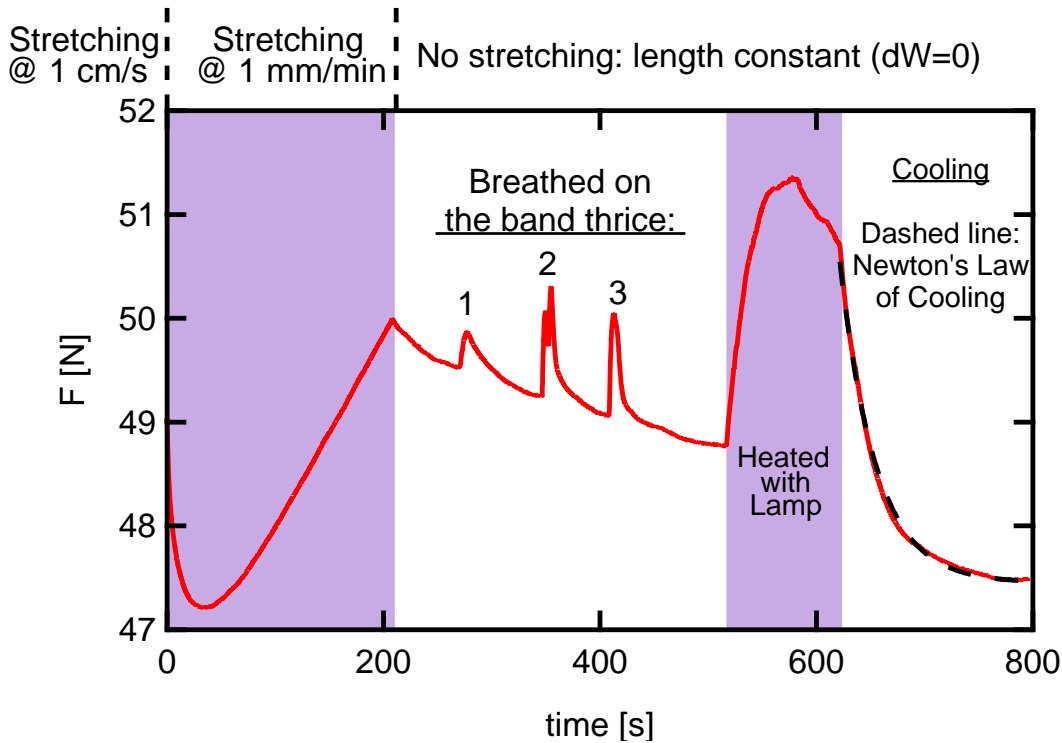}
\caption{A typical plot of tension vs time as measured on the MTS, labeled to indicate what was done to the rubber band. After an initial period of quasistatic extension, the band was breathed on 3 times: once with pursed lips (cool breath, labeled 1), then twice with an open mouth (warm breath, labeled 2 and 3). The initial, quasistatic behavior can only be measured using the MTS, not the double lever.}
\label{fig:sandbox}
\end{figure*}

Starting with the rubber in the unstretched position, the band is adiabatically stretched till the tension reaches a preset tension (our students used 49~N) before data recording starts. An example of the data students will obtain is presented in Fig.~\ref{fig:sandbox}. After the rapid stretching the rubber is quasistatically stretched (at a rate of 1~mm/min) until the tension has increased an additional 1~N. During this step, students using the MTS can directly measure the extension dependence of the tension, as shown in the first 200~seconds of data plotted in Fig.~\ref{fig:sandbox} . Initially, the band cools, as it is still warm from adiabatic stretching, so even as the extension increases, the tension decreases. After approximately 100~seconds, however, the tension approaches linear behavior, as predicted by Eq.~(\ref{eq:avgF}). Discussing this behavior with students is a good way to remind them of the relation they've derived between temperature and force.

Once the tension increases by the additional 1~N, the extension is held constant, and students are asked to breath on the rubber, first with pursed lips (for cool air), then open-mouthed (for warm air), waiting for the rubber to return to room temperature after each breath. This provides another opportunity to teach students about adiabatic processes and call back the prior measurements: blowing with pursed lips generates breath cooler than open-mouthed breath because as the air passes the students' lips, it rapidly (adiabatically) expands.

Next, students are provided with a pistol-grip halogen spotlight, and are instructed to use it to slowly and evenly heat the entire area of the band.  It is important not to overheat a single spot or the band will snap. Once the students have completed these tasks, they are encouraged to explore how changing their protocol for heating the rubber changes its response.  While this portion of the lab is open-ended, students should be reminded to make a note of what they do to the band and when: the data they will use in their reports will contain only the tension vs time. They must provide information about what was done to the band and when.

\subsubsection{Double Lever}
Students are instructed to move the rubber to the stretched position and wait for it to return to room temperature, then re-zero the load cell and start recording the output from the load cell. From this point, students follow the same procedure as described for constant-extension on the MTS above: Students are asked to breath on the rubber, then heat it with a pistol-grip halogen spotlight, and finally to do some open-ended experimentation. A video demonstration of the procedure with synchronized tension data is provided in supplemental materials~\cite{SuppMat}. Again, students should be reminded to make a note of what they do to the band and when.


\subsubsection{Analysis and discussion.}
Because of the open ended nature of the procedure, the analysis undertaken by students is largely qualitative in nature. They should be able to explain the general shape of the peaks in the tension, and contrast peaks corresponding to different methods of heating the band.  Some students may even heat the band to a point that the rate of heat loss to the environment matches the rate at which the lamp heats the band, resulting in a plateau in the tension.

Students are again able to compare their observations to their analytic predictions. Because the band is held at constant extension, $\left(\frac{\partial S}{\partial L}\right)_{U}$ remains approximately constant, so only changes in $T$ will produce changes in $F$. While students are not able to precisely measure temperature dynamically in this excercise, they can predict the way in which the temperature will change as a function of time. The solution to Newton's Law of Cooling predicts that an object, starting at temperature $T_o$, cools to the ambient temperature, $T_a$ , exponentially as $T\left(t\right)=\left(T_{o}-T_{a}\right)e^{-at}$, where $a$ has units of inverse time, and is a material property of the system. An exponential fit to the final cooling period is plotted as a dashed line in Fig.~\ref{fig:sandbox}. The good agreement between the fit and and the data supports the validity of Eq.~(\ref{eq:avgF}).
 
\section{Student Challenges}

Several aspects of this lab challenge students' experimental skills, largely because the structure and premise of the exercise represents a departure from those of the introductory laboratory exercises to which students are accustomed. These challenges not only provide useful learning experiences, but also serve to challenge student preconceptions about experimental goals. One of the most common such challenges arises during the adiabatic stretching portion of the exercise, in large part because students are expected to be critical of their theoretical predictions. Students often misattribute the disparity between the expected and measured values for $\Delta T$ to mistakes made during the lab, or to limitations of the experimental apparatus (e.g. the practical constraints on extension rate). Though students can recite the 'hypothesize, test, refine' model of the Scientific Method by rote, few students are quick to accept that the theory they are testing is flawed. The default assumption is that the theory is right, and the experiment is wrong. We believe this is a consequence of the fact that so many introductory laboratory exercises serve simply to confirm theory, or to measure a physical constant based on the assumed validity of a theory.  For many students, this exercise provides the first opportunity to question the assumptions implicit in their analytic work as a direct response to contradictory experimental evidence.

The second portion of the lab requires careful note-taking and involves relatively unguided analysis. Many students will mislabel details of their plots (such as when they remove the heat source from the rubber), or find that they can't explain features which they understood during the execution of the lab due to lapses in record-keeping. This challenge is also one of the strengths of the exercise: Introductory laboratory instruction often reads like a recipe, yet experimental research requires creativity and thoughtful analysis.  Exercises like this one, which rely on students to determine which features are of physical interest, and to be able to discuss the connection to theory, provide students with important insight into how experimental research is conducted, build experimental skills like note-taking, and serve as a stepping stone for students who wish to pursue further studies in experimental physics. 

Finally, depending on the age and equilibrium length of the band, it may undergo some irreversible stress relaxation during the second part of the exercise. This feature can be observed in Fig.~\ref{fig:sandbox} between 200 and 500s, where there is an overall decreasing trend in the tension. If the band is heated enough, as was done in Fig.~\ref{fig:sandbox} during the period where the band was heated with a lamp, this relaxation is accelerated. Students often identify this feature as 'melting'. Note that after this accelerated relaxation, once the band cools again the relaxation stops. This behavior can prompt interesting discussions about phase transitions, hysteresis and annealing.   

\section{Conclusions}

We find that the physics of rubber are a useful tool to increase student interest in thermodynamics, to extend their experimental skills, and to test their familiarity with the key concepts of thermodynamics.  The adiabatic response of rubber and the Gough-Joule effect challenge student intuition, yet both phenomena are easily explained with a simple 1-D model for a polymer.  We describe an experiment in which students test this simple model against the behavior of real rubber. This instructional lab activity has been a part of our undergraduate Thermal Physics course for second- and third-year students for the past two years.

In the first part of the experiment, students quantitatively compare theoretical predictions against the measured behavior of adiabatically stretched rubber.  The comparison between experiment and theory requires students to explicitly calculate the conversion of configurational entropy to thermal kinetic entropy. The experimental observation of this process demonstrates that entropy is a measureable state variable. At the same time, discrepancies between theoretical prediction and experimental results are used to motivate students to examine the assumptions made in their analytic work and hypothesize what changes might produce a more accurate model. Students then measure the response of rubber to perturbations in temperature while at constant extension.  This open-ended exercise qualitatively matches the behavior predicted by the students' simple analytic model, and challenges students to come up with their own experimental or analytic approaches to testing their model.  One example is by demonstrating that the tension of a cooling rubber band is proportional to Newton's Law of Cooling.

The experiment is an excellent example of the capacity of thermodynamics to predict the properties of everyday objects, and of the capacity of experiment to inform theory. It leads students to explore the First and Second Laws of Thermodynamics, to engage with the concepts of entropy, work and temperature, to understand adiabatic processes, and build their comfort level working with simple mechanistic models. The exercise is also a powerful example of how microscopic physics can manifest in macroscopic systems: while the analysis is based on the behavior of molecules measured in angstroms, the students observe the effects in a piece of rubber measured in centimeters. For instructors, it adds a valuable example to their arsenal, alongside the Ising Model and Ideal Gas. While the hysteretic nature of rubber, and its tendency to anneal at relatively low temperatures leads to some undesirable practical considerations, such as sample degradation and confusing contributions to experimental results, even these provide potentially valuable teaching opportunities.  This experiment has been a valuable component of our undergraduate thermodynamics curriculum, and has improved the engagement and interest of students in the course.
\begin{acknowledgements}
This work was supported by the NCSU Office of Postdoctoral Affairs under the 2013 Professional Development Award, and by the National Science Foundation under Grant No. DMR-1206808. 
\end{acknowledgements}
\bibliographystyle{ajpdjg}
\bibliography{My_Collection}



\end{document}